\documentclass[aps,prl,twocolumn,a4paper]{revtex4-1}
\usepackage[dvipdfmx]{graphicx}

\usepackage{amsmath,amsthm,amssymb}
\usepackage{amsmath,bm}
\usepackage{color}
\usepackage{ifthen}

\usepackage{multirow,bigdelim}

\newcount \commentout
\newcommand{\1}{\mbox{1}\hspace{-0.25em}\mbox{l}}

\newcommand{\mytitle}{
Reduction of one-dimensional non-Hermitian point-gap topology by correlations
}

\newlength{\figwidth}
\newlength{\figlarge}
\begin{document}
\title{
\mytitle
}
\author{
Tsuneya Yoshida
}
\author{
Yasuhiro Hatsugai
}
\affiliation{
Department of Physics, University of Tsukuba, Ibaraki 305-8571, Japan
}
\date{\today}
\begin{abstract}
In spite of extensive works on the non-Hermitian topology, correlations effects remain crucial questions.
We hereby analyze correlated non-Hermitian systems with special emphasis on the one-dimensional point-gap topology.
Specifically, our analysis elucidates that correlations result in reduction of the topological classification $\mathbb{Z}\times \mathbb{Z} \to \mathbb{Z}$ for systems of one synthetic dimension with charge $\mathrm{U(1)}$ symmetry and spin-parity symmetry. 
Furthermore, we analyze an extended Hatano-Nelson chain which exhibits striking correlation effects; correlations destroy the skin effect at the non-interacting level.
This fragility of the skin effect against interactions is consistent with the reduction of the point-gap topology in the one spatial dimension.
The above discoveries shed new light on the topology of correlated systems and open up new directions of researches on non-Hermitian topological physics.
\end{abstract}
\maketitle


\textbf{
Introduction--.
}
Topological insulators and superconductors have been extensively analyzed in these 15 years~\cite{Thouless_PRL1982,Hatsugai_PRL93,Kitaev_chain_01,Kane_2DZ2_PRL05,Kane_Z2TI_PRL05_2,Qi_TFT_PRB08,TI_review_Hasan10,TI_review_Qi10,Sato_JPSJ16}.
In particular, considerable efforts have been devoted to understanding correlation effects on the non-trivial topology, which has revealed a variety of unique phenomena.
For instance, correlation effects induce topological ordered phases~\cite{Tsui_FQHEExp_PRL82,Laughlin_FQHE_PRL83,Jain_FQHE_PRL89,Wen_Textbook04,Levin_LevinWenModel_PRB05,Kitaev_ToricCode_AnnPhys06,Kitaev_KitaevHoney_AnnPhys06} which host anyons.
In addition, it has turned out that correlation effects change $\mathbb{Z}$-classification of topological superconductors at the mean-field level to $\mathbb{Z}_8$-classification~\cite{Z_to_Zn_Fidkowski_PRB10}. 
Such a reduction phenomenon of possible topological phases for a given symmetry class has been theoretically reported for arbitrary spatial dimensions~\cite{Turner_ZtoZ8_PRB11,Fidkowski_1Dclassificatin_PRB11,YaoRyu_Z_to_Z8_2013,Ryu_Z_to_Z8_2013,Qi_Z_to_Z8_2013,LuVishwanath_gauge2013,Levin2013,Isobe_Fu2015,Superlattice_Yoshida17,Fidkowski_Z162013,Wang_Potter_Senthil2014,Metlitski_3Dinteraction2014,Wang_Senthil2014,You_Cenke2014,Morimoto_2015}.
Furthermore, a theoretical work~\cite{CMJian_ZtoZnSynthetic_PRX18} has elucidated that the reduction can also occur in synthetic dimensions which are considered to be fabricated in cold atoms~\cite{Boada_SynthDimExp_PRL12,Celi_SynthDimExp_PRL14,Nakajima_TopoPump_NatPhys16,Lohse_TopoPump_NatPhys16}.
These developments reveal the ubiquity of the reduction phenomena.

Along with the above significant progress, understanding of the non-Hermitian band topology has been rapidly developed in these years~\cite{Hatano_PRL96,CMBender_PRL98,Hu_nH_PRB11,Esaki_nH_PRB11,Bergholtz_Review19,Ashida_nHReview_AdvPhys20}.
Remarkably, it has been elucidated that the point-gap topology induces novel phenomena which do not have Hermitian counterparts~\cite{Alvarez_nHSkin_PRB18,KFlore_nHSkin_PRL18,SYao_nHSkin-1D_PRL18,SYao_nHSkin-2D_PRL18,Yokomizo_BBC_PRL19,EEdvardsson_PRBnHSkinHOTI_PRB19,Gong_class_PRX18,Carlstrom_nHknot_PRA18,Carlstrom_nHknot_PRB19,Zhou_gapped_class_PRB19,Kawabata_gapped_PRX19}.
A prime example is the emergence of the exceptional points~\cite{TKato_EP_book1966,Rotter_EP_JPA09,Berry_EP_CzeJPhys2004,Heiss_EP_JPA12,HShen2017_non-Hermi} (and their symmetry-protected variants~\cite{Budich_SPERs_PRB19,Yoshida_SPERs_PRB19,Okugawa_SPERs_PRB19,Zhou_SPERs_Optica19,Kawabata_gapless_PRL19,Kimura_SPERs_PRB19,Yoshida_nHReview_PTEP20,Delplace_Resul_PRL21,Mandal_EP3_PRL21}) on which the point-gap topology induces band touching for both the real and the imaginary parts.
Another remarkable phenomenon is a non-Hermitian skin effect which results in extreme sensitivity to the presence/absence of boundaries~\cite{SYao_nHSkin-1D_PRL18,Lee_Skin19,Borgnia_ptGapPRL2020,Zhang_BECskin19,Okuma_BECskin19,Yoshida_MSkinPRR20,Okugawa_HOSkin_PRB20,Kawabata_HOSkin_PRB20}.
So far, the non-Hermitian topological band theory has been applied to a wide range of systems from quantum~\cite{Jin_qEP_PRA09,TELeePRL16_Half_quantized,Jose_DissSuperEP_SciRep16,YXuPRL17_exceptional_ring,Lei_qEP_PRA19,VKozii_nH_arXiv17,Zyuzin_nHEP_PRB18,Yoshida_EP_DMFT_PRB18,HShen2018quantum_osci,Papaji_nHEP_PRB19,Matsushita_ER_PRB19,Michishita_EP_PRL20} to classical systems~\cite{Guo_nHExp_PRL09,Ruter_nHExp_NatPhys10,Regensburger_nHExp_Nat12,Zhen_AcciEP_Nat15,Hassan_EP_PRL17,Zhou_FermiArcPH_Science18,Takata_pSSH_PRL18,Ozawa_TopoPhoto_RMP19,Xiao_nHSkin_Exp_NatPhys19,Weidemann_FunnelingnHSkin_Science2020,Hofmann_ExpRecipSkin_19,Helbig_ExpSkin_19,Yoshida_SPERs_mech19,Ghatak_Mech_nHskin_PNAS20,Scheibner_nHmech_PRL20}.

While most of the studies have focused on the non-interacting cases so far, correlation effects on the non-Hermitian topology attract growing interests~\cite{Yoshida_nHFQH19,Xi_nHSPT_SciBull19,Yoshida_nHFQHJ_PRR20,Guo_nHToric_PRB20,Matsumoto_nHToric_PRL20,Zhang_nHToric_Natcomm20,Guo_nHToric_EPL2020,Shackleton_nHFracton_PRR20,Yang_EPKitaev_PRL21,Zhang_nHTMI_PRB20,Liu_nHTMI_RPB20,Xu_nHBM_PRB20,Pan_PTHubb_oQS_PRA20,Mu_MbdySkin_PRB20,Lee_MbdySkin_PRB21,Zhang_CorrSkin_arXiv22,Kawabata_CorrSkin_PRB22,Tsubota_CorrInv_arXiv21,Qin_CorrPolarons_arXiv22,Orito_CorrSkin_PRB22} due to the potential presence of novel non-Hermitian phenomena.
Such interest of correlation effects on the non-Hermitian topology is further enhanced by recent development of technology in cold atoms which allows us to experimentally tune both dissipation and two-body interactions~\cite{Tomita_Zeno_SciAdv17,Takasu_nHPTcoldAtom_PTEP2020}.
Despite these efforts, current understanding of the point-gap topology in correlated systems is quite limited. 
In particular, the knowledge about the reduction of the point-gap topology is limited only to zero dimension~\cite{Yoshida_PtGpZtoZ2_PRB21}, which poses the following significant question: \textit{fate of the higher dimensional point-gap topology under correlations}.

We hereby tackle this question with particular focus on the one-dimensional point-gap topology in both cases of synthetic and spatial dimensions.
We start with the topology in one synthetic dimension. Our analysis reveals the reduction of $\mathbb{Z}\times \mathbb{Z} \to \mathbb{Z}$ for systems with charge $\mathrm{U}(1)$ symmetry and spin-parity symmetry. We end up this conclusion by analyzing a toy model, as well as by an argument in terms of topological invariants.
Furthermore, we analyze an extended Hatano-Nelson chain where such reduction results in a striking phenomenon: fragility of a skin effect against correlations in one spatial dimension.

\textbf{
Topological invariants in one synthetic dimension--.
}
Firstly, we provide a generic argument in terms of topological invariants.
Consider a quantum dot whose many-body Hamiltonian reads
\begin{eqnarray}
\label{eq: gen H mbdy}
\hat{H}&=& \hat{H}_0(\theta)+\hat{H}_{\mathrm{int}},
\end{eqnarray}
%
with 
$
\hat{H}_0(\theta) = \sum_{\alpha \beta} \hat{\Psi}^\dagger_{\alpha} h_{\alpha\beta}(\theta) \hat{\Psi}_{\beta},
$ 
and 
$
\hat{\Psi}^T = (\hat{c}_{a\uparrow},\hat{c}_{a\downarrow},\hat{c}_{b\uparrow},\hat{c}_{b\downarrow},\ldots).
$
The second term $\hat{H}_{\mathrm{int}}$ denotes two-body interactions of fermions. 
Here, one-body Hamiltonian $h(\theta)$ is non-Hermitian and satisfies $h(2\pi)=h(0)$.
The synthetic dimension is parameterized by $\theta$ which corresponds to a tunable parameter in experiments (e.g., a hopping integral in cold atoms).
The operator $\hat{c}^\dagger_{l\sigma}$ ($\hat{c}_{l\sigma}$) creates (annihilates) a fermion in orbital $l$ ($l=a,b,\ldots$) and spin state $\sigma$ ($\sigma=\uparrow,\downarrow$). The subscript $\alpha$ labels the set of $l$ and $\sigma$.

Throughout this paper, we suppose that the Hamiltonian~(\ref{eq: gen H mbdy}) respects the charge $\mathrm{U}(1)$ symmetry and spin-parity symmetry. 
Namely, the zero-dimensional Hamiltonian satisfies
\begin{eqnarray}
\label{eq: U1 symm and R symm}
{}[\hat{H},\hat{N}]=0, \quad  \quad {}[\hat{H},e^{i\pi \hat{S}^z}]= 0, 
\end{eqnarray}
with $\hat{N}=\sum_{\alpha} \hat{\Psi}^\dagger_{\alpha}\hat{\Psi}_{\alpha}$ and $\hat{S}^z=\sum_{l=a,b,\ldots}(\hat{c}^\dagger_{l\uparrow}\hat{c}_{l\uparrow}-\hat{c}^\dagger_{l\downarrow}\hat{c}_{l\downarrow})/2$.

Here, let us discuss the point-gap topology of the above system.
In terms of the one-body Hamiltonian, we can introduce two distinct $\mathbb{Z}$-invariants.
Because the one-body Hamiltonian $h(\theta)$ is periodic in $\theta$, we can introduce the winding number $w$
\begin{eqnarray}
\label{eq: w 1bdy}
w&=& \int^{2\pi}_0 \!\! \frac{d\theta}{2\pi i} \partial_{\theta} \mathrm{tr} \log \left[ h(\theta)-\epsilon_{\mathrm{ref}}\1 \right],
\end{eqnarray}
with the reference energy $\epsilon_{\mathrm{ref}}\in \mathbb{C}$.
The derivative with respect to $\theta$ is denoted by $\partial_{\theta}$.
The symbol ``$\mathrm{tr}$" denotes the trace of a matrix (i.e., $\mathrm{tr} h =\sum_{\alpha} h_{\alpha\alpha}$).

In addition, we can introduce spin winding number $w_{\mathrm{s}}$
\begin{eqnarray}
\label{eq: ws 1bdy}
w_{\mathrm{s}}&=& \int^{2\pi}_0 \!\! \frac{d\theta}{4\pi i} \partial_{\theta} \mathrm{tr}\left[ s^z \log \left( h(\theta)-\epsilon_{\mathrm{ref}}\1\right)\right],
\end{eqnarray}
with $(s^z)_{\alpha\beta}= \mathrm{sgn}(\sigma)\delta_{\alpha\beta}$. Here, $\delta_{\alpha\beta}$ takes $1$ ($0$) for $\alpha=\beta$ ($\alpha\neq \beta$), and $\mathrm{sgn}(\sigma)$ takes $1$ ($-1$) for $\sigma=\uparrow$ $(\downarrow)$.
For the spin winding number the spin-parity symmetry is essential; the one-body Hamiltonian satisfies
$
[s^z,h(\theta)] = 0
$
in the presence of the spin-parity symmetry.

The above results indicates that in the presence of the $\mathrm{U}(1)$ symmetry and the spin-parity symmetry, the point-gap topology of $h(\theta)$ is characterized by two distinct $\mathbb{Z}$-invariants.

Now, let us discuss the point-gap topology of the many-body Hamiltonian. 
In the presence of the spin-parity symmetry, the Hamiltonian $\hat{H}$ can be block-diagonalized with $\hat{N}$ and $\hat{P}:=(-1)^{\hat{N}_{\uparrow}}=e^{i\frac{\pi}{2}\hat{N}}e^{i\pi \hat{S}^z}$.
Here, $\hat{N}_{\uparrow}$ denotes the operator of total number of fermions in the up-spin state.
Therefore, for each Fock space, the following many-body winding number $W_{(N,P)}$ can be introduced~\cite{sector_ftnt};
\begin{eqnarray}
\label{eq: W Mbdy}
W_{(N,P)}(E_{\mathrm{ref}})&=& \int^{2\pi}_0 \!\! \frac{d\theta}{2\pi i} \partial_{\theta} \mathrm{Tr} \log [ \hat{H}_{(N,P)} -E_{\mathrm{ref}}\1],
\end{eqnarray}
where $N$ and $P$ are eigenvalues of $\hat{N}$ and $\hat{P}$, respectively. 
The reference energy is denoted by $E_{\mathrm{ref}}\in \mathbb{C}$.
By $\hat{H}_{(N,P)}$, we denote the many-body Hamiltonian for the subsector with $(N,P)$.
The symbol ``$\mathrm{Tr}$" denotes the trace over the subsector of the Fock space.

In the absence of interactions, eigenvalues of the many-body Hamiltonian $\hat{H}_{(N,P)}$ for each Fock space is computed from the eigenvalues of the one-body Hamiltonian $h(\theta)$ whose point-gap topology is characterized by $w$ and $w_{\mathrm{s}}$.

The above results indicate that the point-gap topology of the one-body Hamiltonian $h(\theta)$ is characterized by a set of two $\mathbb{Z}$-invariants $(w,w_{\mathrm{s}})$ while the topology of the many-body Hamiltonian $\hat{H}$ is characterized by the $\mathbb{Z}$-invariant $W_{(N,P)}$ for each sector of the Fock space.
This fact implies that the non-trivial topology characterized by $(w,w_{\mathrm{s}})=(0,1)$ is trivialized by introducing the interactions, which we see below.

\textbf{
Two orbital quantum dot: non-interacting case--.
}
As a specific case of Eq.~(\ref{eq: gen H mbdy}), let us consider a two-orbital quantum dot ($l=a,b$) with a diagonal matrix $h(\theta)$ 
[$h_{\alpha\beta}(\theta)=h_{\alpha}(\theta)\delta_{\alpha\beta}$] whose diagonal elements are written as
\begin{eqnarray}
h_\alpha(\theta)&=&  \lambda e^{i\theta} \delta_{\alpha,(a,\uparrow)} +  \lambda e^{-i\theta}\delta_{\alpha,(a,\downarrow)} +i\epsilon_{l\sigma}\delta_{\alpha,(l,\sigma)}.
\end{eqnarray}
Here, $\lambda$ and $\epsilon_{l\sigma}$ ($l=a,b$ and $\sigma=\uparrow,\downarrow$) are real numbers.
At the non-interacting level, couplings between orbitals are absent. 
The one-body Hamiltonian of orbital $a$ corresponds to the small cycle limit of an extended Hatano-Nelson chain under the twisted boundary condition [see Eq.~(\ref{eq: 1bdyHami_eHN})].

The topology of $h(\theta)$ is characterized as $(w,w_{\mathrm{s}})=(0,1)$ for $\epsilon_{\mathrm{ref}}=0$ and $|\epsilon_{a\sigma}|<\lambda$ ($\sigma=\uparrow,\downarrow$).
\begin{figure}[!h]
\begin{minipage}{1\hsize}
\begin{center}
\includegraphics[width=1\hsize,clip]{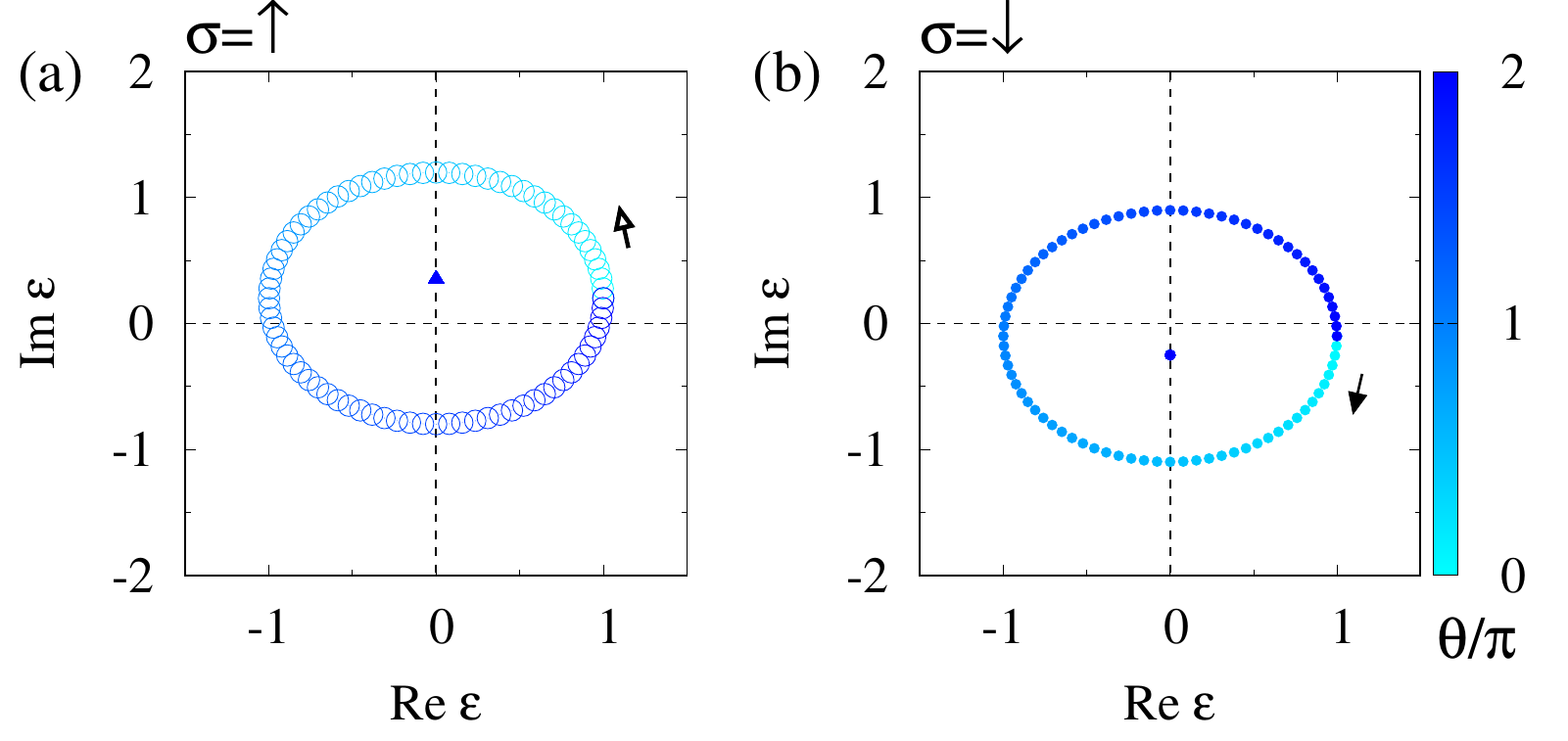}
\end{center}
\end{minipage}

\caption{
Spectral flow of the one-body Hamiltonian $h(\theta)$. Data for the subsector with $\sigma=\uparrow$ [$\sigma=\downarrow$] are plotted in panel (a) [(b)]. The color denotes the value of $\theta$. 
The data are obtained for $(\lambda,\epsilon_{a\uparrow},\epsilon_{a\downarrow},\epsilon_{b\uparrow},\epsilon_{b\downarrow})=(1,0.2,-0.1,0.35,-0.25)$.
}
\label{fig: wcws}
\end{figure}
To be concrete, we plot a spectral flow of the one-body Hamiltonian in Fig.~\ref{fig: wcws} for $(\lambda, \epsilon_{a\uparrow},\epsilon_{a\downarrow},\epsilon_{b\uparrow},\epsilon_{b\downarrow})=(1, 0.2,-0.1,0.35,-0.25)$.
This figure indicate that increasing $\theta$ from $0$ to $2\pi$, an eigenvalue winds around the origin in the clockwise (counter-clockwise) direction for the subsector $\sigma=\uparrow$ ($\sigma=\downarrow$).
The above numerical data support that the topology of $h(\theta)$ is characterized as $(w,w_{\mathrm{s}})=(0,1)$.

\begin{figure}[!h]
\begin{minipage}{1\hsize}
\begin{center}
\includegraphics[width=1\hsize,clip]{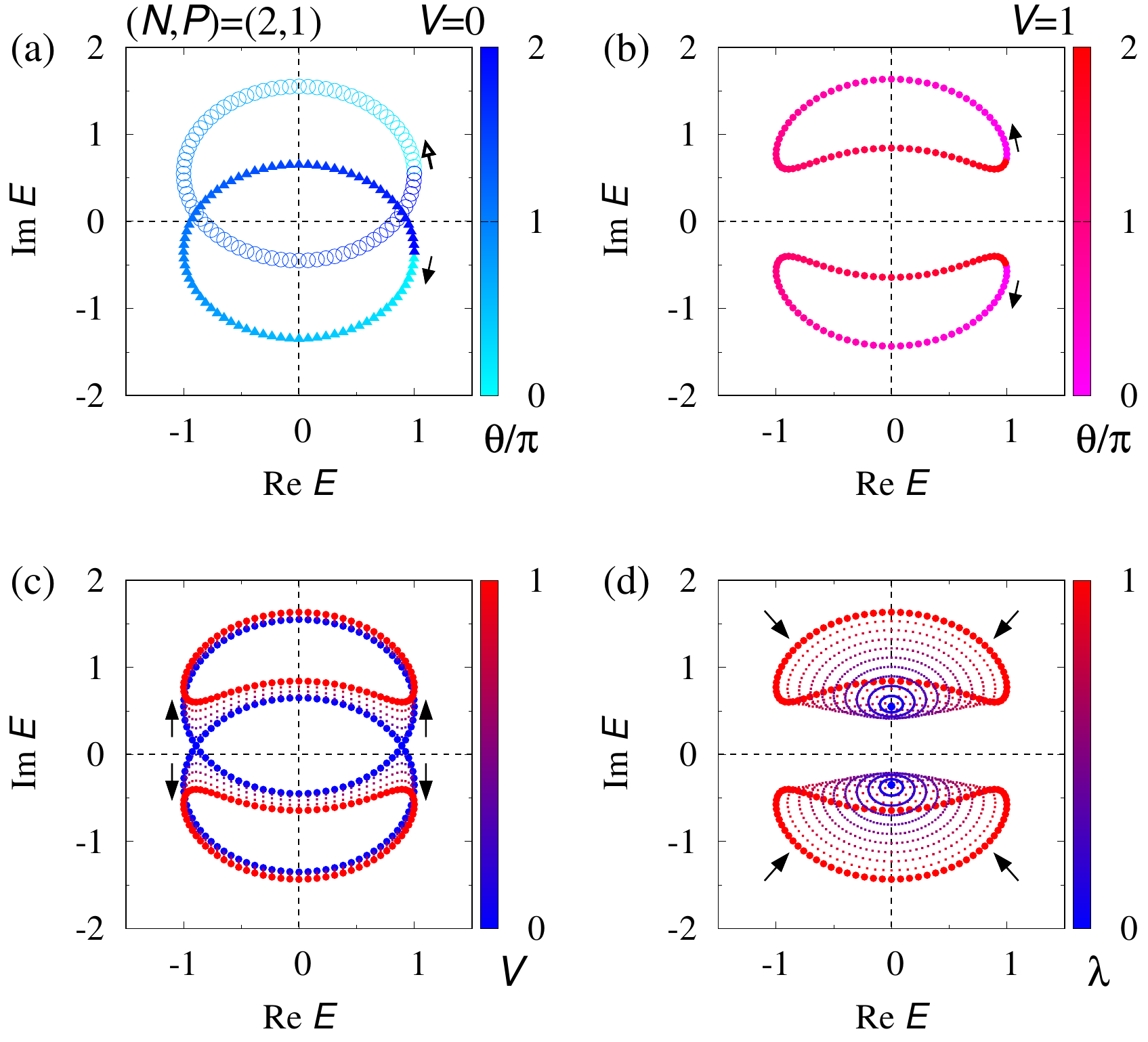}
\end{center}
\end{minipage}
\caption{
Spectral flow of the many-body Hamiltonian for the subsector with $(N,P)=(2,1)$.
(a): Spectral flow for $J=V=0$ and $\lambda=1$. 
Data for the subsector with $(N,S^z)=(2,1)$ [$(2,-1)$] are plotted with open circles [closed triangles].
(b): Spectral flow for $V=J=\lambda=1$. 
In panels (a)~and~(b), we can see that the eigenvalues flow as denoted by arrows with increasing $\theta$ from $0$ to $2\pi$.
(c): Spectral flow for several values of $V$ ($J=V$) at $\lambda=1$. 
With increasing $V$ from $0$ to $1$, the eigenvalues flow as denoted by arrows.
(d): Spectral flow for several values of $\lambda$ for $V=J=\sqrt{\lambda}$. 
With decreasing $\lambda$ from $1$ to $0$, the eigenvalues flow as denoted by arrows.
The data are obtained for $(\epsilon_{a\uparrow},\epsilon_{a\downarrow},\epsilon_{b\uparrow},\epsilon_{b\downarrow})=(0.2,-0.1,0.35,-0.25)$.
}
\label{fig: MbdyE N2P1}
\end{figure}

Figure~\ref{fig: MbdyE N2P1}(a) displays a spectral flow of the many-body Hamiltonian $\hat{H}_0$ for the subsector with $(N,P)=(2,1)$ of the Fock space.
We can observe the loop structure of the spectral flow due to the topology of the one-body Hamiltonian $h(\theta)$.
However, this figure indicates that $\hat{H}_{(2,1)}$ is topologically trivial [i.e., $W_{(2,1)}=0$] for $E_{\mathrm{ref}}=0$ because an eigenvalue winds around the origin in the clockwise direction, and the other eigenvalue winds around the origin in the opposite direction.

\textbf{
Two orbital quantum dot: interacting case--.
}
Now, let us introduce the following two-body interaction
\begin{eqnarray}
\label{eq: JSS VSS}
\hat{H}_{\mathrm{int}} &=& \frac{iJ}{2}(\hat{S}^+_{a}\hat{S}^-_{b}+\mathrm{h.c.}) +\frac{iV}{2}(\hat{S}^+_{a}\hat{S}^+_{b}+\mathrm{h.c.}), 
\end{eqnarray}
with real numbers $J$ and $V$.
Here, ``$\mathrm{h.c.}$" denotes the Hermitian conjugate of the corresponding operator [e.g., $iJ(\hat{S}^+_{a}\hat{S}^-_{b}+\mathrm{h.c.})=iJ(\hat{S}^+_{a}\hat{S}^-_{b}+\hat{S}^-_{a}\hat{S}^+_{b})$].
The spin operator $\hat{S}^\pm_{l}$ is defined as $\hat{S}^\pm_{l}= \hat{S}^x_{l}\pm i \hat{S}^y_{l}$ with $\hat{S}^{x(y)}_{l}$ being the $x$ ($y$) component of the spin operator for orbital $l$.
The above two-body interactions respect charge $\mathrm{U}(1)$ symmetry and spin-parity symmetry; applying the operator $e^{i\pi\hat{S}^z}$ transforms the spin operators as $e^{i\pi \hat{S}^z} \hat{S}^\pm_l e^{-i\pi \hat{S}^z}=-\hat{S}^\pm_l$, meaning that the interactions respect spin-parity symmetry.

For the sake of simplicity, we focus on the subsector with $(N,P)=(2,1)$.
The results for the subsector $(N,P)=(2,-1)$ are provided in Sec.~\ref{sec: 0D (N,P)=(2,-1) app} of Supplemental Material~\cite{supple}.
Figure~\ref{fig: MbdyE N2P1}(b) displays the spectral flow for $V=J=1$.
Remarkably, this figure indicate that the interactions open a imaginary gap; interactions split the loops which wind the origin at the non-interacting level [see Fig.~\ref{fig: MbdyE N2P1}(a)].

This fact indicates that interactions [Eq.~(\ref{eq: JSS VSS})] allow a smooth deformation of the spectral flow for $\lambda=1$ to that for $\lambda=0$ without closing the point-gap at $E_{\mathrm{ref}}=0$ the latter of which is obviously  trivial.
Indeed, the following deformation smoothly connects the Hamiltonian $\hat{H}(\theta)$ for $\lambda=1$ and that for $\lambda=0$:
(i) Increasing $V$ from $0$ to $1$ for $\lambda=1$ and $J=V$ [see Fig.~\ref{fig: MbdyE N2P1}(c)]; 
(ii) Decreasing $\lambda$ from $1$ to $0$ for $J=V=\sqrt{\lambda}$ [see Fig.~\ref{fig: MbdyE N2P1}(d)].
This deformation demonstrates that the many-body Hamiltonian $\hat{H}_{(2,1)}(\theta)$ is topologically trivial.

We note that difference of the symmetry constraint of the spin-parity symmetry is essential for the imaginary gap at $\mathrm{Im}E=0$ in Fig.~\ref{fig: MbdyE N2P1}(b).
As discussed above, the symmetry constraints~(\ref{eq: U1 symm and R symm}), which results in $[s^z, h(\theta)]=0$, forbids hybridization terms between two distinct subsectors with $(N,S^z)$.
In contrast, the symmetry constraint allows such hybridization terms of two-body interactions $\hat{H}_{\mathrm{int}}$.
Therefore, the two-body interactions can destroy the loop structure arising from the non-trivial topology of the one-body Hamiltonian  [see Figs.~\ref{fig: MbdyE N2P1}(a)~and~\ref{fig: MbdyE N2P1}(b)].

For instance, in the subsector with $(N,P)=(2,1)$, the Hamiltonian is written as
\begin{eqnarray}
\hat{H}_{(2,1)}&=& 
\left(
\begin{array}{cc}
\lambda e^{i\theta}+i\epsilon_{a\uparrow}+i\epsilon_{b\uparrow} & \frac{iV}{2} \\
\frac{iV}{2} & \lambda e^{-i\theta}+i\epsilon_{a\downarrow}+i\epsilon_{b\downarrow}
\end{array}
\right). \nonumber 
\end{eqnarray}
Here, we have chosen the following basis vectors spanning the subsector of the Fock space
$
\left(
\hat{c}^\dagger_{a\uparrow}\hat{c}^\dagger_{b\uparrow}|0\rangle,\
\hat{c}^\dagger_{a\downarrow}\hat{c}^\dagger_{b\downarrow}|0\rangle
\right)
$.
The vacuum state is denoted by $|0\rangle$ (i.e., $\hat{c}_{l\sigma}|0\rangle=0$ for arbitrary $l$ and $\sigma$).

Diagonalizing the above Hamiltonian, we obtain
\begin{eqnarray}
\label{eq: E (NP)=(2,1)}
E_{\pm}&=& \lambda \cos\theta+i\delta_0 \pm i\sqrt{(\sin\theta+\delta_3)^2+(\frac{V}{2})^2},
\end{eqnarray}
with $2\delta_0=\epsilon_{a\uparrow}+\epsilon_{b\uparrow}+\epsilon_{a\downarrow}+\epsilon_{b\downarrow}$ and $2\delta_3=\epsilon_{a\uparrow}+\epsilon_{b\uparrow}-\epsilon_{a\downarrow}-\epsilon_{b\downarrow}$.
Equation~(\ref{eq: E (NP)=(2,1)}) elucidates that spin-parity symmetry allows the hybridization term between states with $(N,S^z)=(2,1)$ and $(N,S^z)=(2,-1)$ which opens the line-gap $\mathrm{Im}[E_+(\theta)-E_-(\theta)]>0$ [see Fig.~\ref{fig: MbdyE N2P1}(b)].
In contrast, spin-parity symmetry forbids such hybridization terms for the quadratic Hamiltonian $\hat{H}_0$.

The above numerical results supports that the many-body Hamiltonian $\hat{H}_{(2,1)}(\theta)$ is topologically trivial despite the loop structure due to the topology of the one-body Hamiltonian with $(w,w_{\mathrm{s}})=(0,1)$.

Putting the argument in terms of the topological invariants and the above results of the toy model together, we end up with the reduction of the point-gap topology $\mathbb{Z}\times\mathbb{Z} \to \mathbb{Z}$.

\begin{figure}[!h]
\begin{minipage}{1\hsize}
\begin{center}
\includegraphics[width=1\hsize,clip]{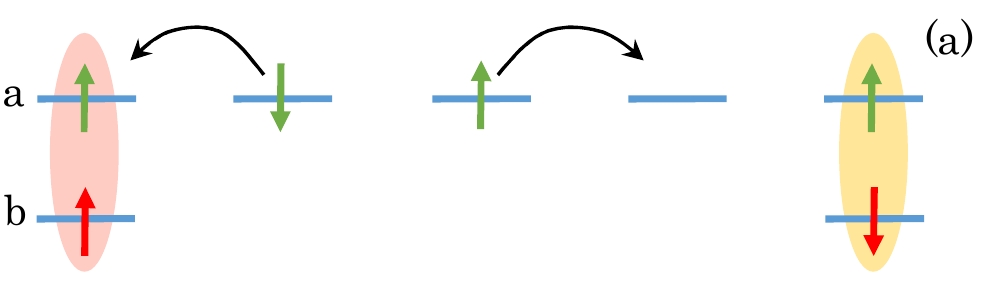}
\end{center}
\end{minipage}
\begin{minipage}{1\hsize}
\begin{center}
\includegraphics[width=1\hsize,clip]{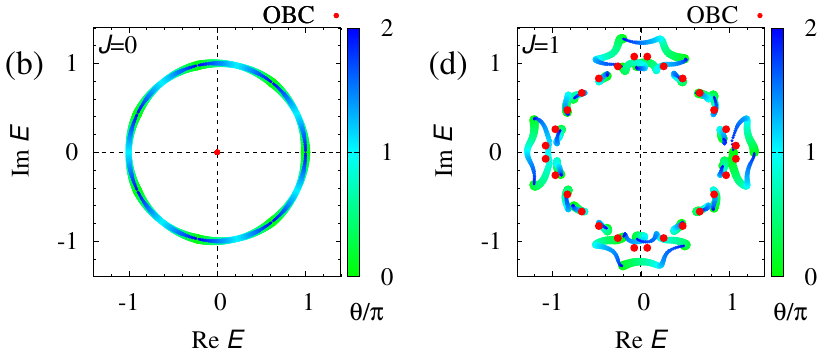}
\end{center}
\end{minipage}
\begin{minipage}{0.48\hsize}
\begin{center}
\includegraphics[width=1\hsize,clip]{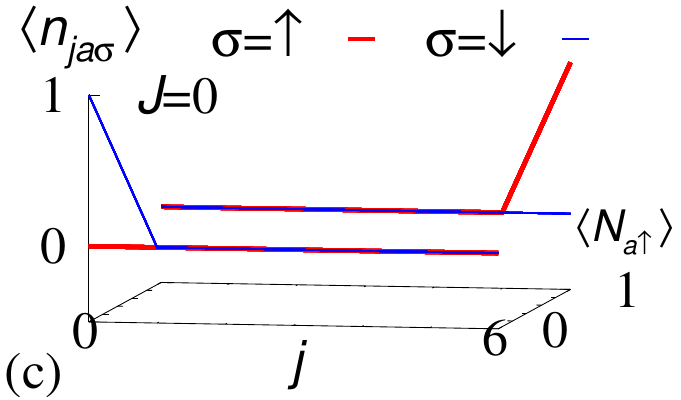}
\end{center}
\end{minipage}
\begin{minipage}{0.48\hsize}
\begin{center}
\includegraphics[width=1\hsize,clip]{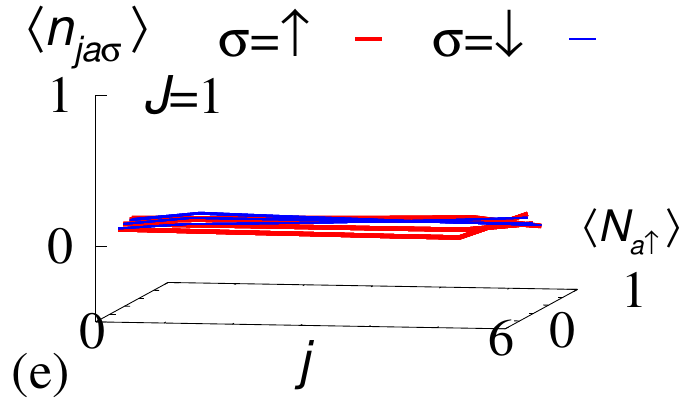}
\end{center}
\end{minipage}
\caption{
(a): Sketch of the extended Hatano-Nelson chain.
(b) [(d)]: Spectral flow for $J=V=0$ [$J=V=1$]. 
Red dots denote the data obtained under the open boundary condition.
(c) [(e)]: Expectation values of $\hat{n}_{ja\sigma}$ for $J=V=0$ [$J=V=1$] under the open boundary condition.
In panels~(c)~and~(e), $\langle \hat{n}_{ja\sigma}\rangle={}_{\mathrm R}\langle \Phi_n| \hat{n}_{ja\sigma} |\Phi_n\rangle_{\mathrm R}$ is plotted against $j$ and $\langle \hat{N}_{a\uparrow}\rangle ={}_{\mathrm R}\langle \Phi_n| \hat{N}_{a\uparrow} |\Phi_n\rangle_{\mathrm R} $.
Here, $|\Phi_n\rangle_{\mathrm{R}}$ ($n=0,1,\ldots$) denote right eigenstates of $\hat{H}_{\mathrm{eHN}}$.
Red (blue) lines denote the data for $\sigma=\uparrow$ ($\sigma=\downarrow$).
These data are obtained for the subsector $(N,P)=(3,-1)$ and a parameter set $(L,t)=(7,1)$.
}
\label{fig: HN_Na1}
\end{figure}

\textbf{
Topology in one spatial dimension and fragility of a skin effect--.
}
By analyzing an extended Hatano-Nelson chain [see Fig.~\ref{fig: HN_Na1}(a)], we elucidate that correlations reduce the point-gap topology in one spatial dimension as is the case in one synthetic dimension. Remarkably, this reduction phenomenon results in fragility of a skin effect against interactions.

Let us consider an extended Hatano-Nelson chain [see Fig.~\ref{fig: HN_Na1}(a)] whose Hamiltonian reads
\begin{subequations}
\begin{eqnarray}
\hat{H}_{\mathrm{eHN}}(\theta)&=& \hat{H}_0(\theta) + \hat{H}_{\mathrm{int}}, \\
\hat{H}_0(\theta)&=& \sum_{k} \hat{\Psi}^\dagger_{k\alpha} h_{\alpha\beta}(k,\theta) \hat{\Psi}_{k\beta},
\end{eqnarray}
\begin{eqnarray}
\hat{H}_{\mathrm{int}}&=& \sum_{j=0,L-1}\left[ J( \hat{S}^+_{ja}\hat{S}^-_{jb}+\mathrm{h.c.}) +iV ( \hat{S}^+_{ja}\hat{S}^+_{jb}+\mathrm{h.c.}) \right],\nonumber \\
\end{eqnarray}
\end{subequations}
with a diagonal matrix $h(k,\theta)$ [$h_{\alpha\beta}(k,\theta)=h_{\alpha}(k,\theta)\delta_{\alpha\beta}$, $(kL/2\pi=0,1,\ldots,L-1)$] whose diagonal elements are 
\begin{eqnarray}
\label{eq: 1bdyHami_eHN}
h_{\alpha}(k,\theta) &=& t\delta_{\alpha,(a,\uparrow)}e^{i(k+\theta/L)} +t\delta_{\alpha,(a,\downarrow)}e^{-i(k+\theta/L)}.
\end{eqnarray}
Here, we have imposed the twisted boundary condition in order to compute the winding numbers (for more details, see Sec.~\ref{sec: eHN defs app} of Supplemental Material~\cite{supple}).
The operator $\hat{\Psi}_{k\alpha}$ is the Fourier transformed annihilation operator $\Psi_{k\alpha}:=\frac{1}{\sqrt{L}}\sum_{j=0,\ldots,L-1}e^{ikj} \Psi_{j\alpha}$ with $\Psi_{j}^T=(\hat{c}_{ja\uparrow},\hat{c}_{ja\downarrow},\hat{c}_{jb\uparrow},\hat{c}_{jb\downarrow})$. 
The two-body term $\hat{H}_{\mathrm{int}}$ describes the interaction between fermions in orbital $a$ and localized fermions in orbital $b$.
This model also preserves charge $\mathrm{U(1)}$ and spin-parity symmetry, meaning that $\hat{H}_{\mathrm{eHN}}(\theta)$ can be block-diagonalized with $\hat{N}$ and $\hat{P}=(-1)^{\hat{N}_{a\uparrow}+\hat{N}_{b\uparrow}}$ where $\hat{N}_{l\sigma}$ and $\hat{N}$ are defined as $\hat{N}_{l\sigma}=\sum_{j}\hat{c}^\dagger_{jl\sigma}\hat{c}_{jl\sigma}$ and $\hat{N}=\sum_{l\sigma}\hat{N}_{l\sigma}$, respectively.
The Hamiltonian $\hat{H}_{\mathrm{eHN}}(\theta)$ also commutes with $\hat{n}_{jb}=\sum_{\sigma} \hat{c}^\dagger_{jb\sigma} \hat{c}_{jb\sigma}$ for $j=0, \ L-1$, and thus, we suppose that orbital $b$ is occupied at both edges ($j=0, \ L-1$).

Now, we demonstrate that for $N_a=1$ (i.e., $N=3$), a skin effect observed at the non-interacting level is fragile against the two-body interactions due to trivial topology of the many-body Hamiltonian.
Let us start with the non-interacting level. 
Under the twisted boundary condition, the spectral flow shows a loop structure [see Fig.~\ref{fig: HN_Na1}(b)] due to the point-gap topology of one-body Hamiltonian $h(\theta):=\bigoplus_{k}h(k,\theta)$ characterized by $(w,w_{\mathrm{s}})=(0,1)$ for $\epsilon_{\mathrm{ref}}=0$.
This non-trivial topology of $h$ induces the skin effect at the non-interacting level.
In the presence of the boundaries, all of the eigenvalues [$E_n$ ($n=0,1,\ldots$)] become zero in contrast to the eigenvalues in the absence of the boundaries [see Fig.~\ref{fig: HN_Na1}(b)]. 
In addition, a fermion in the up- (down-) spin state is localized around the right (left) edge under the open boundary condition [see Fig.~\ref{fig: HN_Na1}(c)]. 

\begin{figure}[!h]
\begin{minipage}{0.48\hsize}
\begin{center}
\includegraphics[width=1\hsize,clip]{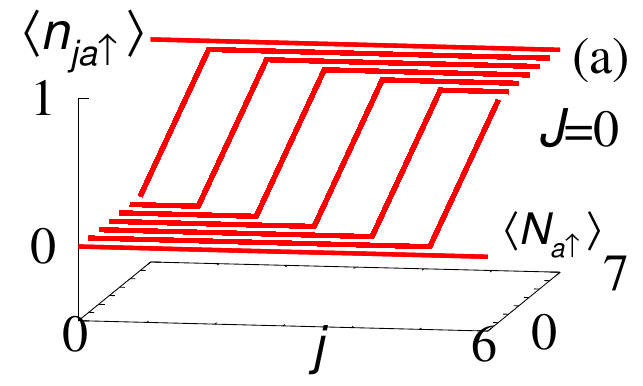}
\end{center}
\end{minipage}
\begin{minipage}{0.48\hsize}
\begin{center}
\includegraphics[width=1\hsize,clip]{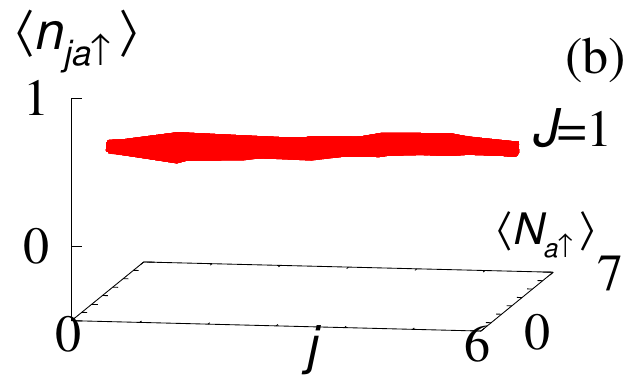}
\end{center}
\end{minipage}
\caption{
Expectation values of $\hat{n}_{ja\uparrow}$ for the subsector $(N,P)=(9,-1)$ and a parameter set $(L,t)=(7,1)$.
In panels (a) and (b), data for $J=V=0$ and $J=V=1$ are plotted, respectively.
These figures are plotted in the same way as Figs.~\ref{fig: HN_Na1}(c)~and~\ref{fig: HN_Na1}(e).
}
\label{fig: HN_NaL}
\end{figure}
However, interactions destroy the above skin effect, which is due to the trivial topology of the many-body Hamiltonian $W_{(3,-1)}=0$ for $E_{\mathrm{ref}}=0$ (for computation of the many-body winding number, see Fig.~\ref{fig: HN_N3P-1_app} of Supplemental Material~\cite{supple}). 
Because of the trivial topology, we can observe that interactions destroy the loop structure of the spectral flow and open a line gap for $J=V=1$ [see Fig.~\ref{fig: HN_Na1}(d)], which is also confirmed by analysis based on the perturbation theory (see Sec.~\ref{sec: eHN ptb app} of Supplemental Material~\cite{supple}). 
This result verifies the reduction of the point-gap topology $\mathbb{Z}\times \mathbb{Z} \to \mathbb{Z}$ for the subsector with $(N,P)=(3,-1)$.
Correspondingly, the interactions destroy the extreme sensitivity of the spectrum to the presence/absence of boundaries [see Fig.~\ref{fig: HN_Na1}(d)]. 
Furthermore, in the presence of interactions fermions extend to the bulk even under the open boundary condition [see Fig.~\ref{fig: HN_Na1}(e)].
This result is also intuitively understood as follows: 
while the one-body term $\hat{H}_{0}(\theta)$ localizes the fermions in orbital $a$ and the up- (down-) spin state around the right (left) edge, the two-body interactions $\hat{H}_{\mathrm{int}}$ flip their spins at edges, which suppresses the effects of boundaries.
The above results indicate that the skin effect observed at the non-interacting level is fragile against the two-body interactions.
Our numerical calculation indicate that such fragility of the skin effect is also observed for the case of many fermions in orbital $a$ [see Fig.~\ref{fig: HN_NaL}].
More detailed data are provided in Sec.~\ref{sec: eHN numeri app} of Supplemental Material~\cite{supple}.

\textbf{
Summary and discussion--.
}
We have analyzed correlation effects on the one-dimensional point-gap topology in both cases of synthetic and spatial dimensions.
Our analysis has elucidated that the reduction $\mathbb{Z}\times\mathbb{Z}\to\mathbb{Z}$ occurs for systems of synthetic one dimension with charge $\mathrm{U(1)}$ symmetry and spin-parity symmetry. 
This conclusion is obtained by the argument of topological invariants as well as by explicit analysis of the toy model. 
Furthermore, we have also analyzed the extended Hatano-Nelson chain which exhibits striking correlation effects: correlations reduce the point-gap topology and destroy the skin effect at the non-interacting level. 

The above discoveries shed new light on non-Hermitian correlated systems and open up a new directions of researches on non-Hermitian topological physics. 
For instance, the above results imply the possibility of similar reduction phenomena for other cases of symmetry and dimensions. As well as the above theoretical open question, experimental observation of the reduction is also a significant issue to be addressed. We expect that cold atoms are promising candidate where interactions and non-Hermiticity can be tuned in experiments.

\textbf{
Acknowledgements--.
}
This work is also supported by JSPS KAKENHI Grant No.~JP21K13850 and also by JST CREST, Grant No. JPMJCR19T1, Japan.

\begin{thebibliography}{127}
\bibitem [{sec()}]{sector_ftnt}%
  \BibitemOpen
  \href@noop {} {}\bibinfo {note} {We note that introducing additional terms
  makes other subsectors irrelevant. For instance adding the term
  $\mu(\hat{N}-2)$ with $\mu\in \mathbb{C}$ to Eq.~(\ref{eq: gen H mbdy}) can
  make subsectors labeled by $N=0,1,3,4$ irrelevant to the topology for
  $E_{\mathrm{ref}}=0$}\BibitemShut {NoStop}%
\end{thebibliography}
%


\clearpage

\renewcommand{\thesection}{S\arabic{section}}
\renewcommand{\theequation}{S\arabic{equation}}
\setcounter{equation}{0}
\renewcommand{\thefigure}{S\arabic{figure}}
\setcounter{figure}{0}
\renewcommand{\thetable}{S\arabic{table}}
\setcounter{table}{0}
\makeatletter
\c@secnumdepth = 2
\makeatother

\onecolumngrid
\begin{center}
 {\large \textmd{Supplemental Materials:} \\[0.3em]
 {\bfseries 
\mytitle
 }
 }
\end{center}

\setcounter{page}{1}

\section{
Analysis of a two orbital model for the subsector with $(N,P)=(2,-1)$
}
\label{sec: 0D (N,P)=(2,-1) app}
In the main text, we have seen that interactions open a line-gap for the subsector with $(N,P)=(2,1)$, which is consistent with the trivial topology $W_{(2,1)}=0$ for $E_{\mathrm{ref}}=0$. 
In this section, we show that a similar behavior is observed for the subsector with $(N,P)=(2,-1)$.
\begin{figure}[!h]
\begin{minipage}{0.5\hsize}
\begin{center}
\includegraphics[width=1\hsize,clip]{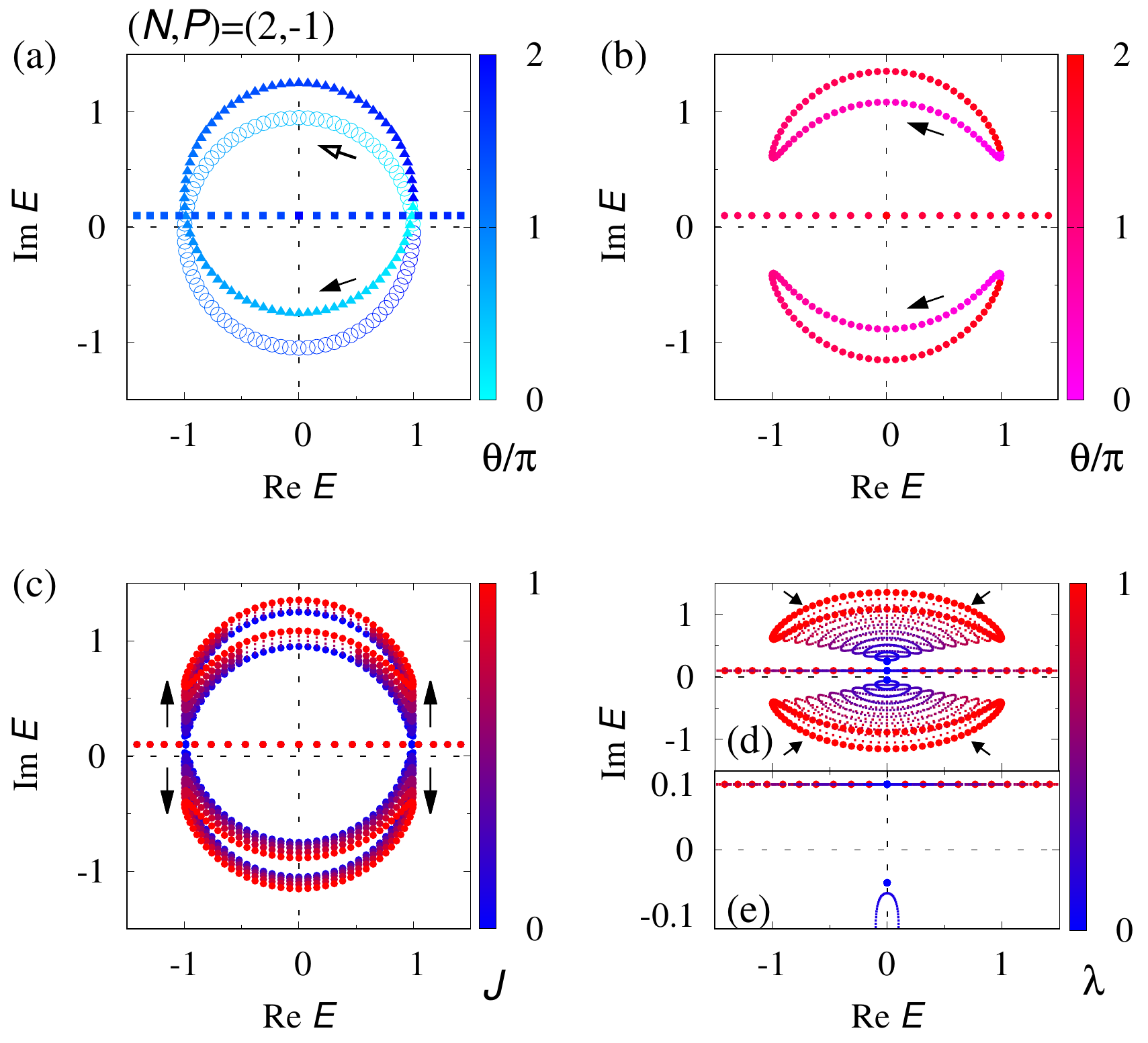}
\end{center}
\end{minipage}
\caption{
Spectral flow of the many-body Hamiltonian for the subsector with $(N,P)=(2,-1)$.
Panels~{(a)-(d)} are plotted in the same way as panels~{(a)-(d)} in Fig.~\ref{fig: MbdyE N2P1}.
Panel~(e) is a magnified version of the range $ -0.12 \leq  \mathrm{Im}E \leq 0.12 $ in panel~(d).
}
\label{fig: MbdyE N2P-1}
\end{figure}

Despite the non-trivial topology of the one-body Hamiltonian, the many-body winding number takes zero [$W_{(2,-1)}=0$] for $E_{\mathrm{ref}}=0$ as shown in Fig.~\ref{fig: MbdyE N2P-1}(a).
Correspondingly, the spectrum of the many-body Hamiltonian $\hat{H}_{(2,-1)}$ can smoothly shrink to the points [see Figs.~\ref{fig: MbdyE N2P-1}(b)-\ref{fig: MbdyE N2P-1}(d)].

In this subsector, the interaction $J$ is essential for destruction of the loop structure observed in Fig.~\ref{fig: MbdyE N2P-1}(a), which can be seen as follows.
In the subsector with $(N,P)=(2,-1)$, the Hamiltonian is written as
\begin{subequations}
\begin{eqnarray}
\hat{H}_{(2,-1)}&=& \hat{H}_{0(2,-1)}+\hat{H}_{\mathrm{int}(2,-1)},
\end{eqnarray}
\begin{eqnarray}
\hat{H}_{0(2,-1)} &=& 
\mathrm{diag}\left(\lambda e^{i\theta}+i\epsilon_{a\uparrow}+i\epsilon_{b\downarrow}, \lambda e^{-i\theta}+i\epsilon_{a\downarrow}+i\epsilon_{b\uparrow} \right. \nonumber \\
&&\left. 2\lambda \cos\theta +i\epsilon_{a\uparrow}+i\epsilon_{a\downarrow}, i\epsilon_{b\uparrow}+i\epsilon_{b\downarrow} \right),
\end{eqnarray}
\begin{eqnarray}
\hat{H}_{\mathrm{int}(2,-1)} &=& 
\frac{iJ}{2}
\left(
\begin{array}{cccc}
0  & 1& 0 & 0  \\
1  & 0& 0 & 0 \\
0  & 0& 0 & 0  \\
0  & 0& 0 & 0 
\end{array}
\right),
\end{eqnarray}
\end{subequations}
with $\mathrm{diag}(\cdots)$ denoting a diagonal matrix.
Here, we have chosen the following basis vectors spanning the subsector of the Fock space
\begin{eqnarray}
\left( 
\hat{c}^\dagger_{a\uparrow}\hat{c}^\dagger_{b\downarrow} |0\rangle,
\hat{c}^\dagger_{a\downarrow}\hat{c}^\dagger_{b\uparrow} |0\rangle,
\hat{c}^\dagger_{a\uparrow}\hat{c}^\dagger_{a\downarrow} |0\rangle,
\hat{c}^\dagger_{b\uparrow}\hat{c}^\dagger_{b\downarrow} |0\rangle
\right).
\end{eqnarray}
Diagonalizing the Hamiltonian, we obtain
\begin{eqnarray}
E_{\pm}&=& \lambda \cos \theta + i\delta'_0 \pm i \sqrt{(\sin\theta+\delta'_3)^2+(\frac{J}{2})^2},  \\
E' &=& 2\lambda\cos\theta +i\epsilon_{a\uparrow}+i\epsilon_{a\downarrow},  \\
E''&=& i\epsilon_{b\uparrow}+i\epsilon_{b\downarrow}, 
\end{eqnarray}
with 
$2\delta'_0= \epsilon_{a\uparrow}+\epsilon_{b\downarrow}+ \epsilon_{a\downarrow}+\epsilon_{b\uparrow}$
and 
$2\delta'_3= \epsilon_{a\uparrow}+\epsilon_{b\downarrow}-(\epsilon_{a\downarrow}+\epsilon_{b\uparrow})$.
The above results elucidate that the spin-parity symmetry allows the two-body interaction which splits loop structure observed in Fig.~\ref{fig: MbdyE N2P-1}(a).

\section{
Details of the extended Hatano-Nelson chain
}
\label{sec: eHN app}

\subsection{
Hamiltonian under the twisted boundary condition
}
\label{sec: eHN defs app}

We provide the explicit form of the extended Hatano-Nelson chain under the twisted boundary condition.
The Hamiltonian reads
\begin{subequations}
\begin{eqnarray}
\hat{H}_{\mathrm{eHN}}&=& \hat{H}_0(\theta) + \hat{H}_{\mathrm{int}},
\end{eqnarray}
\begin{eqnarray}
\label{eq: eHN H0 app}
\hat{H}_{0}(\theta)&=& t \left[ e^{i\theta} \hat{c}^\dagger_{0a\uparrow} \hat{c}_{L-1a\uparrow} +\sum^{L-2}_{j=0} \hat{c}^\dagger_{j+1a\uparrow} \hat{c}_{ja\uparrow} \right] + t\left[ e^{-i\theta} \hat{c}^\dagger_{L-1a\downarrow} \hat{c}_{0a\downarrow} +\sum^{L-1}_{j=1} \hat{c}^\dagger_{j-1a\downarrow} \hat{c}_{ja\downarrow} \right],
\end{eqnarray}
\begin{eqnarray}
\hat{H}_{\mathrm{int}}(\theta)&=& \sum_{j=0,L-1}\left[ \frac{J}{2}( \hat{S}^+_{ja}\hat{S}^-_{jb}+\hat{S}^-_{ja}\hat{S}^+_{jb}) +iV ( \hat{S}^+_{ja}\hat{S}^+_{jb}+\hat{S}^-_{ja}\hat{S}^-_{jb}) \right].
\end{eqnarray}
\end{subequations}

Under a gauge transformation $\hat{c}_{ja\sigma} \to e^{-i\frac{\theta}{L}j} \hat{c}_{ja\sigma}$, the one-body term is written as
\begin{eqnarray}
\hat{H}_{0}(\theta)&=&  \sum^{L-1}_{j=0} \left[ t e^{i \theta /L} \hat{c}^\dagger_{j+1a\uparrow} \hat{c}_{ja\uparrow} + t e^{-i\theta/L} \hat{c}^\dagger_{j-1a\downarrow} \hat{c}_{ja\downarrow} \right], 
\end{eqnarray}
with $\hat{c}^\dagger_{La\uparrow}:=\hat{c}^\dagger_{0a\uparrow}$ and $\hat{c}^\dagger_{-1a\downarrow}:=\hat{c}^\dagger_{L-1a\downarrow}$.

Applying the Fourier transformation to the above Hamiltonian yields Eq.~(\ref{eq: 1bdyHami_eHN}).
We note that under the open boundary condition, hopping terms between sites $j=0$ and $j=L-1$ [i.e., the first and the third terms of Eq.~(\ref{eq: eHN H0 app})] become zero.

In the presence of charge $\mathrm{U(1)}$ symmetry and spin-parity symmetry, the one-body Hamiltonian is characterized by $w$ and $w_{\mathrm{s}}$ [Eqs.~(\ref{eq: w 1bdy})~and~(\ref{eq: ws 1bdy})] with $h(\theta):=\oplus_{k}h(k,\theta)$.
The topology of the many-body Hamiltonian for given subsector with $(N,P)$ is characterized by the many-body winding number [Eq.~(\ref{eq: W Mbdy})] with $\hat{H}_{(N,P)}:=\hat{H}_{\mathrm{eHN}(N,P)}$.
Here, $\hat{H}_{\mathrm{eHN}(N,P)}$ denotes the many-body Hamiltonian of the extended Hatano-Nelson model for the given subsector with $(N,P)$.

\subsection{
Analysis based on the perturbation theory
}
\label{sec: eHN ptb app}

Based on the perturbation theory, we confirm that interactions open a line-gap as shown in Fig.~\ref{fig: HN_Na1}(d).
As mentioned in the main text, we suppose that orbital $b$ is occupied at both edges ($j=0, \ L-1$).

Suppose that interactions are sufficiently weak. 
In the subsector of $(N,P)=(3,-1)$, $\hat{H}_{0}(\theta)$ is written as
\begin{eqnarray}
\label{eq: H0 matrix ptb app}
\hat{H}_{0(3,-1)} &=& 
t\omega^n
\left(
\begin{array}{cccc}
e^{i\frac{\theta}{L}} & 0                     & 0                      & 0 \\
0                     & e^{i\frac{\theta}{L}} & 0                      & 0 \\
0                     & 0                     & e^{-i\frac{\theta}{L}} & 0 \\
0                     & 0                     & 0                      & e^{-i\frac{\theta}{L}}
\end{array}
\right),
\end{eqnarray}
with $\omega=e^{\frac{2\pi i}{L}}$ and 
the basis
\begin{eqnarray}
\label{eq: basis app}
\left(
|n\uparrow;\uparrow \uparrow \rangle, 
|n\uparrow;\downarrow \downarrow \rangle,
|n\downarrow;\uparrow \downarrow \rangle, 
|n\downarrow;\downarrow \uparrow \rangle
\right),
\end{eqnarray}
for given $n$ ($n=0,1,2,\ldots, L-1$).
Here, $|n\sigma;\sigma' \sigma''\rangle$ is defined as $|n\sigma;\sigma' \sigma''\rangle := \bar{\hat{d}}_{n\sigma} \hat{c}^\dagger_{0b\sigma'}  \hat{c}^\dagger_{L-1b\sigma''} |0\rangle$ with 
$\bar{\hat{d}}_{n\uparrow}:=\sum_{j}\hat{c}^\dagger_{ja\uparrow}R_{jn}$ and $\bar{\hat{d}}_{n\downarrow}:=\sum_{j}\hat{c}^\dagger_{ja\downarrow}L^*_{jn}$.
Matrices $R$ and $L^\dagger$ ($R_{jn}:=\frac{1}{\sqrt{L}} \omega^{-nj} $ and $L^\dagger_{nj}:=\frac{1}{\sqrt{L}} \omega^{nj}$) diagonalize the matrix $h$ ($h_{ij}= t \delta_{i,j+1}$)
\begin{eqnarray}
L^\dagger h R&=& t\, \mathrm{diag}(1,\omega,\omega^2,\cdots,\omega^{L-1})
\end{eqnarray}
which corresponds to the kinetic term of fermions in orbital $a$ and the up-spin state for $\theta=0$ [see Eq.~(\ref{eq: eHN H0 app}) and Fig.~\ref{fig: HN_Na1}(a)].
Here $\mathrm{diag}(\cdots)$ describes a diagonal matrix.
Introducing the operators
$\hat{d}_{n\uparrow}:=\sum_{j}\hat{c}_{ja\uparrow}(L^\dagger)_{nj}$ and $\hat{d}_{n\downarrow}:=\sum_{j}\hat{c}_{ja\downarrow}(R^T)_{nj}$,
we have anti-commutation relations
\begin{eqnarray}
{} \{ \hat{d}_{n\sigma},  \bar{\hat{d}}_{m\sigma'}\}&=& \delta_{nm}\delta_{\sigma\sigma'},
\end{eqnarray}
for $n,m=0,1,2,\ldots, L-1$ and $\sigma, \sigma'=\uparrow,\downarrow$, which can be seen by noting the relations $\sum_{j}L^\dagger_{nj}R_{jm}=\delta_{nm}$ and $\{\hat{c}_{il\sigma},\hat{c}^{\dagger}_{jl'\sigma'} \}= \delta_{ij}\delta_{ll'}\delta_{\sigma\sigma'}$.
We note that $(\bar{\hat{d}}_{n\sigma})^\dagger=\hat{d} _{n\sigma}$ holds due to the relation $R^*_{jn}=L^\dagger_{nj}$.

Now, let us compute energy eigenvalues at the first order of the interactions.
Firstly, we note the following relations.
\begin{eqnarray}
\hat{S}^+_{ja} |m\downarrow;\sigma \sigma'\rangle &=& \sum_i \hat{S}^+_{ja} L^*_{im} \hat{c}^\dagger_{ia\downarrow}\hat{c}^\dagger_{0b\sigma}\hat{c}^\dagger_{L-1b\sigma'} | 0\rangle \nonumber \\
                                                  &=& L^*_{jm} \hat{c}^\dagger_{ja\uparrow}\hat{c}^\dagger_{0b\sigma}\hat{c}^\dagger_{L-1b\sigma'} | 0\rangle \nonumber \\
                                                  &=& \sum_{n} L^*_{jm} (L^\dagger)_{nj} \bar{\hat{d}}_{n\uparrow}\hat{c}^\dagger_{0b\sigma}\hat{c}^\dagger_{L-1b\sigma'} | 0\rangle \nonumber \\
                                                  &=& \frac{1}{L}\sum_{n}  \omega^{(n+m)j}|n\uparrow;\sigma\sigma'\rangle,\\
\hat{S}^-_{ja} |m\uparrow;\sigma \sigma'\rangle &=& \sum_i \hat{S}^-_{ja} R_{im} \hat{c}^\dagger_{ia\uparrow}\hat{c}^\dagger_{0b\sigma}\hat{c}^\dagger_{L-1b\sigma'} | 0\rangle \nonumber \\
                                                  &=& R_{jm} \hat{c}^\dagger_{ja\downarrow}\hat{c}^\dagger_{0b\sigma}\hat{c}^\dagger_{L-1b\sigma'} | 0\rangle \nonumber \\
                                                  &=& \sum_{n} R_{jm} (R^T)_{nj} \bar{\hat{d}}_{n\downarrow}\hat{c}^\dagger_{0b\sigma}\hat{c}^\dagger_{L-1b\sigma'} | 0\rangle \nonumber \\
                                                  &=& \frac{1}{L}\sum_{n}  \omega^{-(n+m)j}|n\downarrow;\sigma\sigma'\rangle.
\end{eqnarray}
Here, we have used the relations 
$\sum_{n} R_{jn}L^\dagger_{ni}=\delta_{ij}$, 
$\hat{c}^\dagger_{ia\uparrow}=\sum_{n} L^\dagger_{ni}\bar{\hat{d}}_{n\uparrow}$,
and
$\hat{c}^\dagger_{ia\downarrow}=\sum_{n} R^T_{ni} \bar{\hat{d}}_{n\downarrow}$.

Thus, at the first order, the Hamiltonian is written as $\hat{H}_{\mathrm{eHN}(3,1)}=\hat{H}_{0(3,1)}+\hat{H}_{\mathrm{int}(3,1)}$ 
with $\hat{H}_{0(3,1)}$ in Eq.~(\ref{eq: H0 matrix ptb app}) and
\begin{eqnarray}
\label{eq: Hami 1st order app}
\hat{H}_{\mathrm{int}(3,-1)} &=& 
\frac{V}{L}
\left(
\begin{array}{cccc}
0           & 0 & \omega^{-2n} & 1 \\
0           & 0 & 0            & 0 \\
\omega^{2n} & 0 & 0            & 0 \\
1           & 0 & 0            & 0
\end{array}
\right)
+
\frac{iJ}{L}
\left(
\begin{array}{cccc}
0 & 0           & 0 & 0 \\
0 & 0           & 1 & \omega^{-2n}  \\
0 & 1           & 0 & 0 \\
0 & \omega^{2n} & 0 & 0
\end{array}
\right),
\end{eqnarray}
for the basis defined in Eq.~(\ref{eq: basis app}).

The eigenvalues of $\hat{H}_{\mathrm{eHN}(3,1)}$ are written as
\begin{subequations}
\label{eq: E 1st order app}
\begin{eqnarray}
E_{\mathrm{p},\pm} &=& t \omega^n \left( \cos (\frac{\theta}{L}) \pm \sqrt{C^2_{\mathrm{p}}-\sin^2 (\frac{\theta}{L}) }\right), \\
E_{\mathrm{m},\pm} &=& t \omega^n \left( \cos (\frac{\theta}{L}) \pm \sqrt{C^2_{\mathrm{m}}-\sin^2 (\frac{\theta}{L}) }\right),
\end{eqnarray}
with
\begin{eqnarray}
C^2_{\mathrm{p}} &=& \frac{1}{(t\omega^nL)^2} \left[ (V^2-J^2)+\sqrt{V^4+J^4-2V^2J^2\mathrm{Re}[\omega^{4n}]} \right], \\
C^2_{\mathrm{m}} &=& \frac{1}{(t\omega^nL)^2} \left[ (V^2-J^2)-\sqrt{V^4+J^4-2V^2J^2\mathrm{Re}[\omega^{4n}]} \right].
\end{eqnarray}
\end{subequations}
As well as by directly diagonalizing the matrix, the eigenvalues are obtained by taking square of the matrices (see below).
These results indicate that interactions lift four-fold degeneracy observed for $\theta=0$.
Specifically, the imaginary parts of $C_{\mathrm{p}}$ and $C_{\mathrm{m}}$ lift the degeneracy.
To see this, firstly, let us suppose that the imaginary parts are zero ($\mathrm{Im}C_{\mathrm{p}}=\mathrm{Im}C_{\mathrm{m}}=0$), then, Eq.~(\ref{eq: E 1st order app}a) indicates that exceptional points emerge at certain $\theta$ [i.e., $C^2_{\mathrm{p}}\geq 0$ holds, and $\mathrm{C}^2_{\mathrm{p}} -\sin^2(\theta/L)=0$ can be satisfied].
On the other hand, the finite imaginary parts lift the degeneracy at $\theta=0$ without inducing exceptional points [i.e., $\mathrm{C}^2_{\mathrm{p}} -\sin^2(\theta/L)\neq 0$ for $ 0\leq  \theta <2\pi$].

Equations~(\ref{eq: E 1st order app}c)~and~(\ref{eq: E 1st order app}d) indicate that the imaginary parts of $C_{\mathrm{p}}$ and $C_{\mathrm{m}}$ can be finite for proper choice of $n$, $J$, and $V$.
Therefore, the above result of the first-order perturbation theory indicate that interactions open a line-gap.

We show that eigenvalues~(\ref{eq: E 1st order app}) can be obtained by taking squares of the matrices. 
Consider the following matrix
\begin{eqnarray}
\tilde{H}&=& 
 \left (x \sigma_0\tau_0
 +y \sigma_0\tau_3
 +a \sigma_1\tau_1
 +b \sigma_2\tau_2
 +c \sigma_0\tau_1
 +d \sigma_3\tau_1
 +f \sigma_0\tau_2
 +g \sigma_3\tau_2
\right),
\end{eqnarray}
with complex numbers $x$, $y$, $a$, $b$, $c$, $d$, $f$ and $g$.
Here $\sigma_0$ and $\tau_0$ denote the $2\times 2$-identity matrix. Pauli matrices are denoted by $\sigma_s$ and $\tau_s$ ($s=1,2,3$).
Matrices $\sigma_\mu\tau_\nu$ ($\mu,\nu=0,1,2,3$) denote $4\times4$-matrices. For instance, $\sigma_1\tau_2$ is written as
\begin{eqnarray}
\sigma_1\tau_2&=& 
\left(
\begin{array}{cccc}
 0 & 0 & 0  & -i \\
 0 & 0 & -i & 0  \\
 0 & i & 0  & 0  \\
 i & 0 & 0  & 0
\end{array}
\right).
\end{eqnarray}
For the following parameter set,
\begin{eqnarray}
\label{eq: params app}
\left(
\begin{array}{c}
x  \\
y  \\
a  \\
b  \\
c  \\
d  \\
f  \\
g
\end{array}
\right)
&=& 
\frac{1}{2L}
\left(
\begin{array}{c}
2Lt \omega^n \cos \frac{\theta}{L} \\
2iLt \omega^n \sin \frac{\theta}{L} \\
V+iJ\\
-V+iJ\\
(V+iJ)\mathrm{Re}(\omega^{2n})  \\
(V-iJ)\mathrm{Re}(\omega^{2n})  \\
(V+iJ)\mathrm{Im}(\omega^{2n})  \\
(V-iJ)\mathrm{Im}(\omega^{2n})  
\end{array}
\right),
\end{eqnarray}
$\tilde{H}$ is reduced to the matrix $\hat{H}_{\mathrm{eHN}(3,-1)}$. 

Taking square of this matrix yields
\begin{eqnarray}
&&( \tilde{H}-x\sigma_0\tau_0)^2- (y^2+a^2+b^2+c^2+d^2+f^2+g^2) \nonumber \\
&=&
-2 ab \sigma_3\tau_3
+2 (cd+fg) \sigma_3\tau_0
+2 ac \sigma_1\tau_0
+2 ag \sigma_2\tau_3
+2 bd \sigma_1\tau_3
+2 bf \sigma_2\tau_0.
\end{eqnarray}
Thus, we have
\begin{eqnarray}
&&
\left[( \tilde{H}-x\sigma_0\tau_0)^2- (y^2+a^2+b^2+c^2+d^2+f^2+g^2)^2\right]^2 \nonumber \\ 
&=&
\left[ \{-2 ab \sigma_3\tau_3 +2 (cd+fg) \sigma_3\tau_0 \}+(2ac\sigma_1\tau_0+2bd\sigma_1\tau_3)+(2ag \sigma_2\tau_3+2bf\sigma_2\tau_0)\right]^2 \nonumber \\
&=&
\{ 
-2ab\sigma_3\tau_3
+2(cd+fg)\sigma_3\tau_0
\}^2
+
(
2ac\sigma_1\tau_0
+2bd\sigma_1\tau_3
)^2
(
2ag \sigma_2\tau_3
+2bf\sigma_2\tau_0
)^2
\nonumber \\
&=&
4\{
(ab)^2
+(cd+fg)^2
+(ac)^2
+(bd)^2
+(ag)^2
+(bf)^2
\}
-8ab(cd+fg)\sigma_0\tau_3
+8acbd\sigma_0\tau_3
+8agbf\sigma_0\tau_3
\nonumber\\
&=&
4\{
(ab)^2
+(cd+fg)^2
+(ac)^2
+(bd)^2
+(ag)^2
+(bf)^2
\}.
\end{eqnarray}
Therefore, the eigenvalues are written as
\begin{eqnarray}
E'_{\mathrm{p}\pm}&=& x\pm \sqrt{y^2+C'^2_{\mathrm{p}}}, \\
E'_{\mathrm{m}\pm}&=& x\pm \sqrt{y^2+C'^2_{\mathrm{m}}},
\end{eqnarray}
with 
\begin{eqnarray}
C'^2_{\mathrm{p}} &=& a^2+b^2+c^2+d^2+f^2+g^2 +2\sqrt{ (ab)^2 +(cd+fg)^2 +(ac)^2 +(bd)^2 +(ag)^2 +(bf)^2}, \\
C'^2_{\mathrm{m}} &=& a^2+b^2+c^2+d^2+f^2+g^2 -2\sqrt{ (ab)^2 +(cd+fg)^2 +(ac)^2 +(bd)^2 +(ag)^2 +(bf)^2}. 
\end{eqnarray}
Thus, choosing the parameters as Eq.~(\ref{eq: params app}), we obtain Eq.~(\ref{eq: E 1st order app}).

\subsection{
Numerical results
}
\label{sec: eHN numeri app}

In the main text, we have briefly discussed the extended Hatano-Nelson chain. Here, let us numerically analyze this system in detail.
\begin{figure}[!h]
\begin{minipage}{0.5\hsize}
\begin{center}
\includegraphics[width=1\hsize,clip]{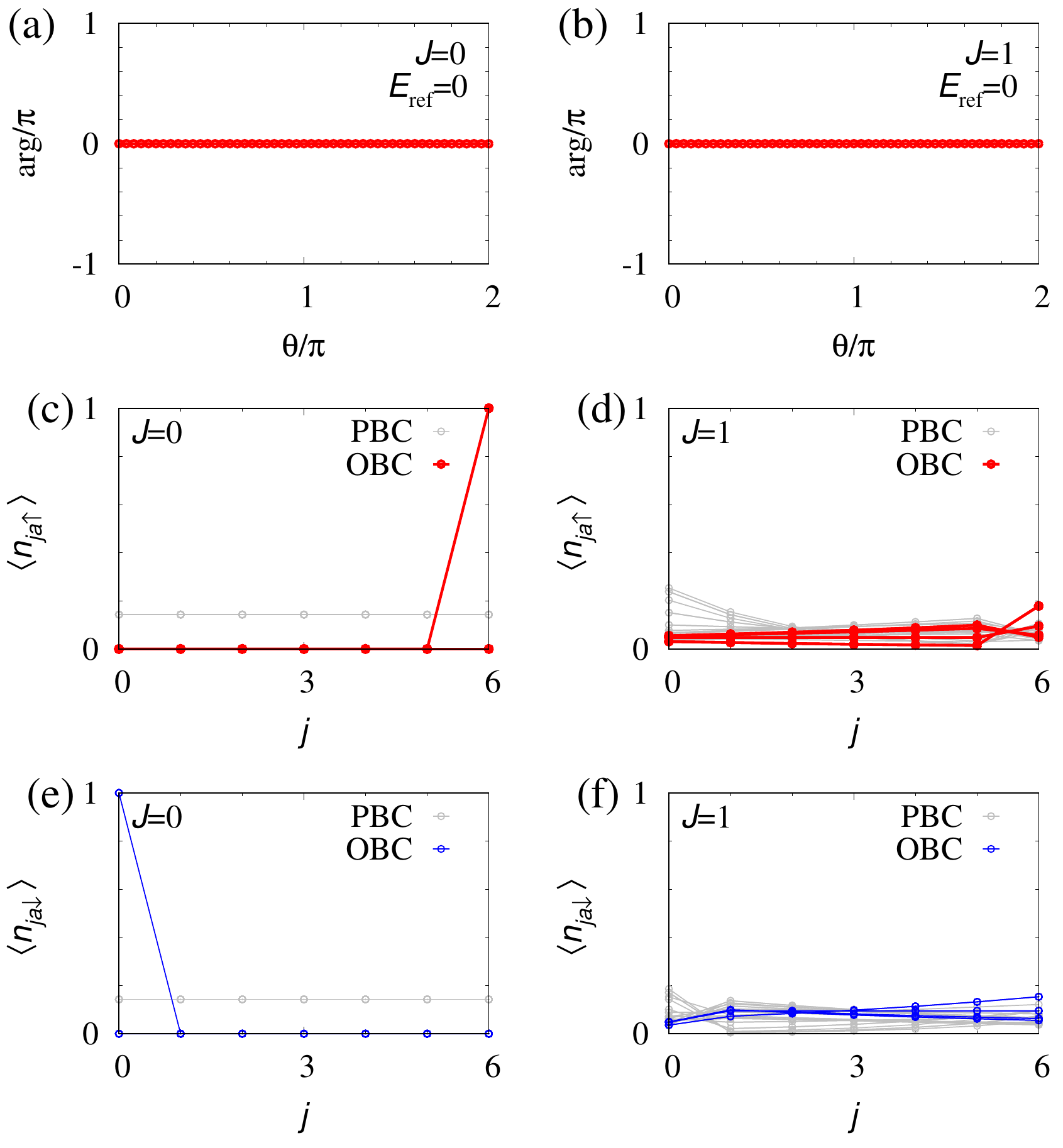}
\end{center}
\end{minipage}
\caption{
Data for $(L,t)=(7,1)$ and the subsector $(N,P)=(3,-1)$.
(a) and (b): The twist angle ($\theta$) dependence of $\mathrm{arg}[\prod_n (E_n-E_{\mathrm{ref}})]$ for $E_{\mathrm{ref}}=0$.
(c) and (d) [(e) and (f)]: Expectation values $\langle \hat{n}_{ja\sigma}\rangle={}_{\mathrm R}\langle \Phi_n| \hat{n}_{ja\sigma} |\Phi_n\rangle_{\mathrm R}$ with $\sigma=\uparrow$ [$\sigma=\downarrow$].
Here, $|\Phi_n\rangle_{\mathrm{R}}$ ($n=0,1,\ldots$) denote right eigenstates of $\hat{H}_{\mathrm{eHN}}$.
Data obtained under the open boundary condition (the periodic boundary condition) are shown with colored (gray) symbols.
Panels~(a),~(c),~and~(e)~[(b),~(d),~and~(f)] display data for $J=V=0$ [$J=V=1$].
}
\label{fig: HN_N3P-1_app}
\end{figure}
Firstly, we focus on the subsector with $(N,P)=(3,-1)$ [see Fig.~\ref{fig: HN_N3P-1_app}].
Although the topology of the one-body Hamiltonian is non-trivial [i.e., $(w,w_{\mathrm{s}})=(0,1)$ for $\epsilon_{\mathrm{ref}}=0$],
the many-body Hamiltonian is topologically trivial [i.e., $W_{(3,-1)}=0$ for $E_{\mathrm{ref}}=0$] as shown in Figs.~\ref{fig: HN_N3P-1_app}(a)~and~\ref{fig: HN_N3P-1_app}(b).
This fact results in the fragility of the skin effect against interactions.
Namely, although the fermion with the up- (down-) spin state is localized at the right (left) edge due to the skin effect in the non-interacting case [see Figs.~\ref{fig: HN_N3P-1_app}(c)~and~\ref{fig: HN_N3P-1_app}(e)], such localization cannot be observed in the presence of the interactions [see Figs.~\ref{fig: HN_N3P-1_app}(d)~and~\ref{fig: HN_N3P-1_app}(f)].
Correspondingly, the extreme sensitivity of the energy spectrum to the boundary condition is not observed for $J=V=1$ [see Figs.~\ref{fig: HN_Na1}(b)~and~\ref{fig: HN_Na1}(d)].

This fragility of the skin effect is intuitively understood as follows: the interactions flip the spin of fermions in orbital $a$, which suppresses the effects of the boundaries.

\begin{figure}[!h]
\begin{minipage}{0.5\hsize}
\begin{center}
\includegraphics[width=1\hsize,clip]{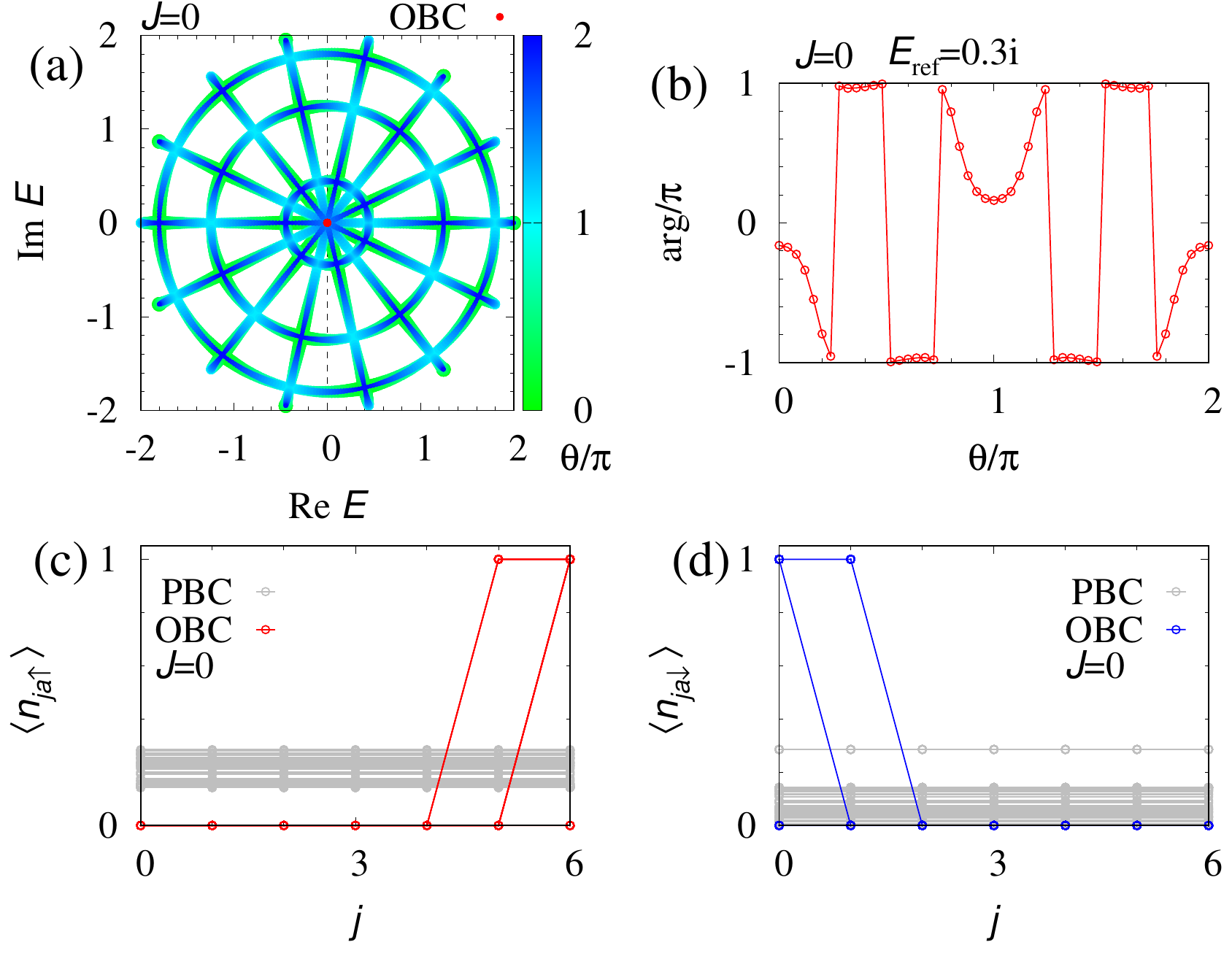}
\end{center}
\end{minipage}
\caption{
Numerical data for $J=V=0$ and the subsector $(N,P)=(4,1)$.
(a): Spectral flow of the many-body Hamiltonian.
(b): The twist angle ($\theta$) dependence of $\mathrm{arg}[\prod_n (E_n-E_{\mathrm{ref}})]$ for $E_{\mathrm{ref}}=0.3i$.
(c) and (d) [(e) and (f)]: Expectation values $\langle \hat{n}_{ja\sigma}\rangle={}_{\mathrm R}\langle \Phi_n| \hat{n}_{ja\sigma} |\Phi_n\rangle_{\mathrm R}$ with $\sigma=\uparrow$ [$\sigma=\downarrow$].
Here, $|\Phi_n\rangle_{\mathrm{R}}$ ($n=0,1,\ldots$) denote right eigenstates of $\hat{H}_{\mathrm{eHN}}(\theta=0)$.
Data obtained under the open boundary condition (the periodic boundary condition) are shown with colored (gray) symbols. These data are obtained for a parameter set $(L,t)=(7,1)$.
}
\label{fig: HN_N4P1_J0_app}
\end{figure}

\begin{figure}[!h]
\begin{minipage}{0.5\hsize}
\begin{center}
\includegraphics[width=1\hsize,clip]{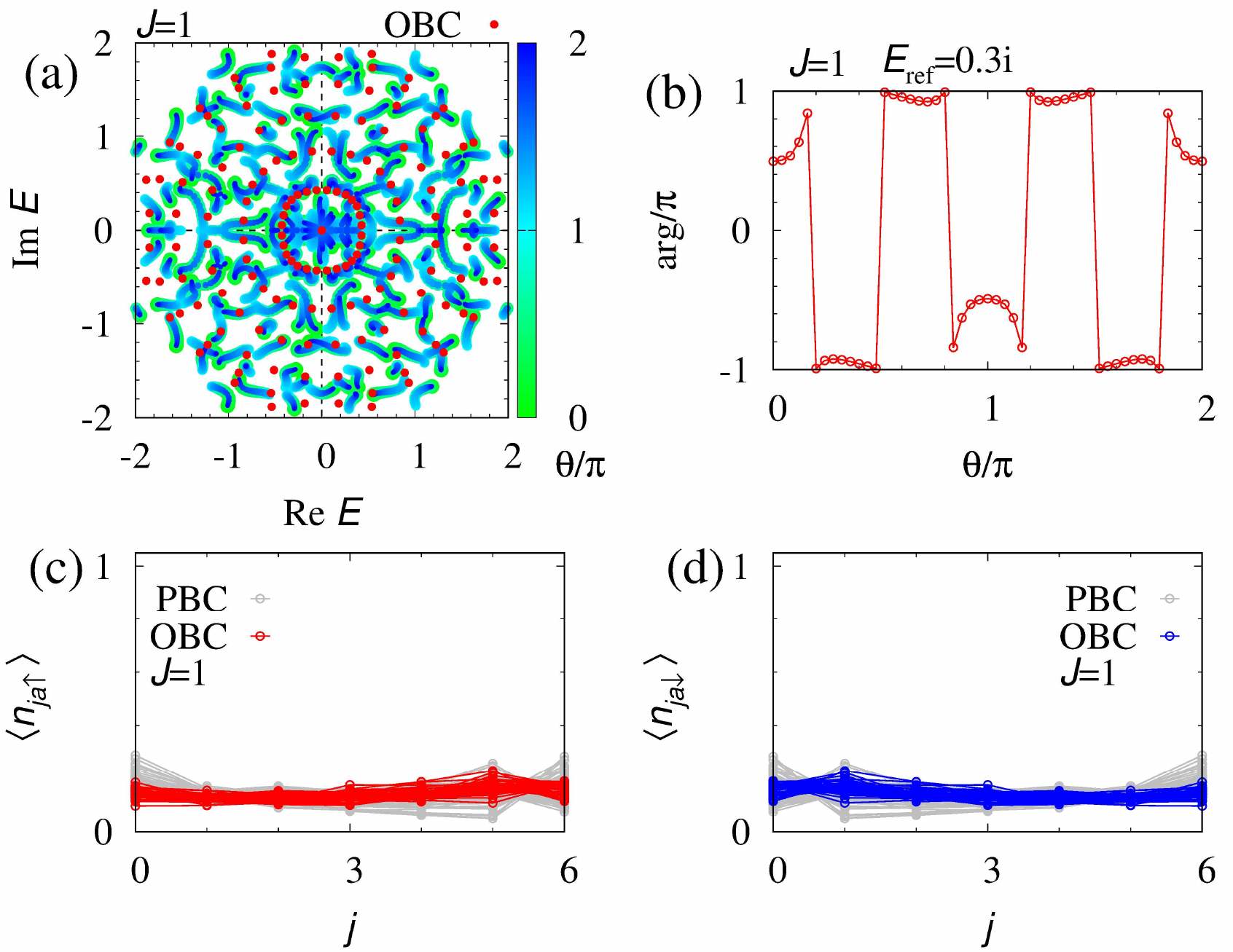}
\end{center}
\end{minipage}
\caption{
Numerical data for $J=V=1$ and the subsector $(N,P)=(4,1)$.
These figures are plotted in the same way as Figs.~\ref{fig: HN_N4P1_J0_app}(a)-\ref{fig: HN_N4P1_J0_app}(d).
}
\label{fig: HN_N4P1_J1_app}
\end{figure}

Now, let us focus on the subsector with $(N,P)=(4,1)$ [see Figs.~\ref{fig: HN_N4P1_J0_app}~and~\ref{fig: HN_N4P1_J1_app}].
Figures~\ref{fig: HN_N4P1_J0_app}(a)~and~\ref{fig: HN_N4P1_J0_app}(b) indicate the topology of the many-body Hamiltonian is trivial.
However, due to the topology of the one-body Hamiltonian, we can observe the extreme sensitivity of the spectrum and expectation values $\langle \hat{n}_{ja\sigma}\rangle $ to the presence/absence of the boundaries [see Figs.~\ref{fig: HN_N4P1_J0_app}(c)~and~\ref{fig: HN_N4P1_J0_app}(d)].
As is the case for $(N,P)=(3,-1)$, such extreme sensitivity is fragile against interactions [see Fig.~\ref{fig: HN_N4P1_J1_app}].

\begin{figure}[!h]
\begin{minipage}{0.5\hsize}
\begin{center}
\includegraphics[width=1\hsize,clip]{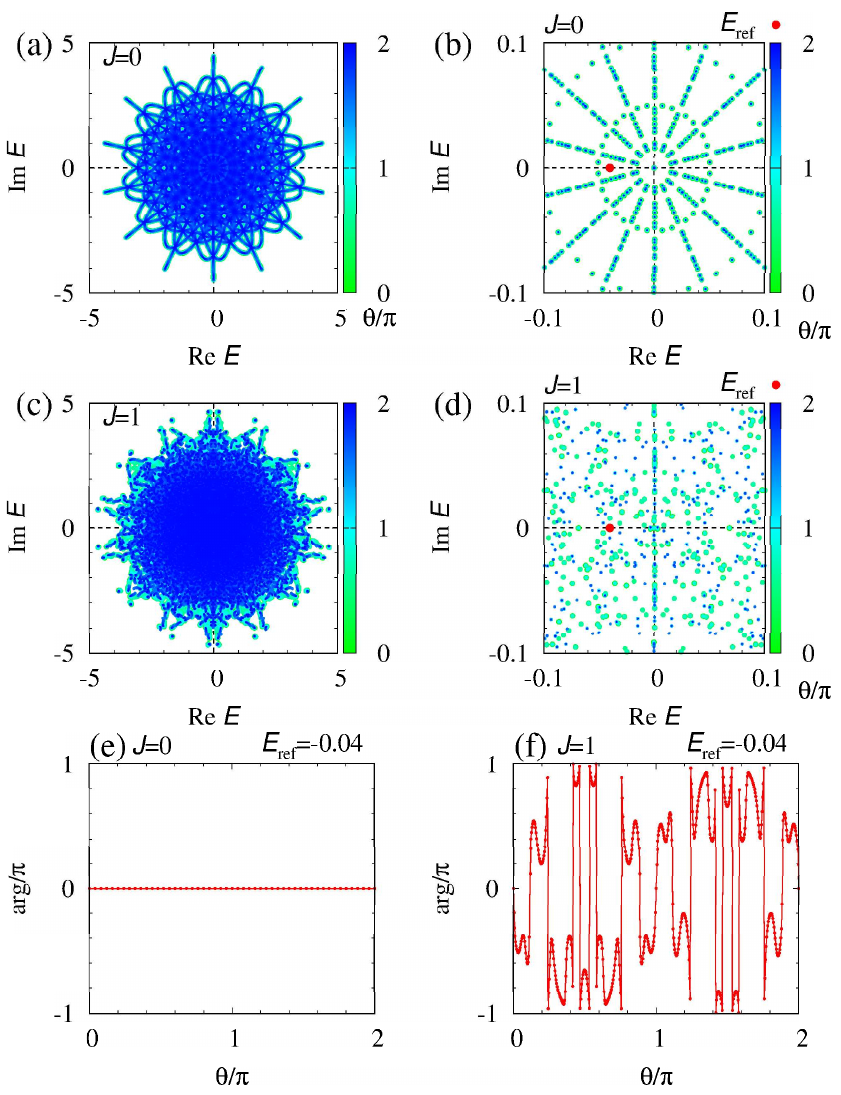}
\end{center}
\end{minipage}
\caption{
Data for $(L,t)=(7,1)$ and the subsector $(N,P)=(9,-1)$.
(a) and (b) [(c) and (d)]: Spectral flow for $J=V=0$ [$J=V=1$]. 
Panel (b) [(d)] is a magnified version of the range $-0.1 \leq \mathrm{Im}E \leq 0.1$ and $-0.1 \leq \mathrm{Re}E \leq 0.1$ in panel (a) [(c)].
(e) and (f): The twist angle ($\theta$) dependence of $\mathrm{arg}[\prod_n (E_n-E_{\mathrm{ref}})]$ for $E_{\mathrm{ref}}=-0.04$ and $J=V=0$ [$J=V=1$].
}
\label{fig: HN_N9P-1_wind}
\end{figure}

\begin{figure}[!h]
\begin{minipage}{1\hsize}
\begin{center}
\includegraphics[width=0.5\hsize,clip]{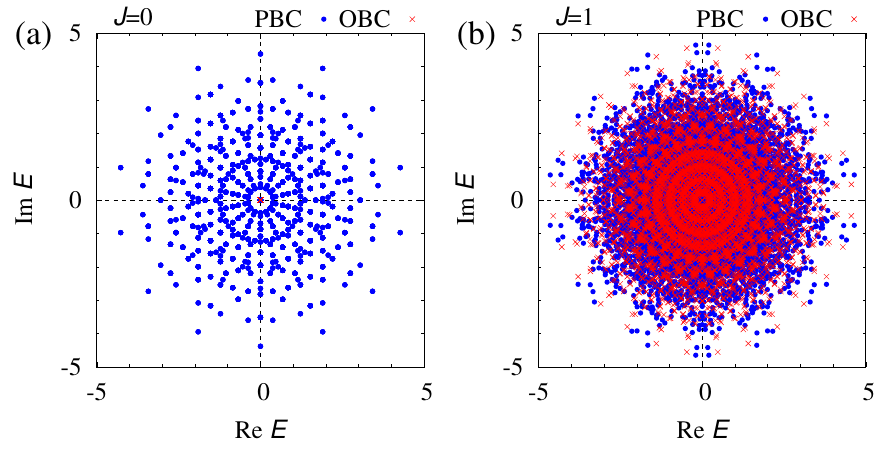}
\end{center}
\end{minipage}
\begin{minipage}{0.48\hsize}
\begin{flushright}
\includegraphics[width=0.5\hsize,clip]{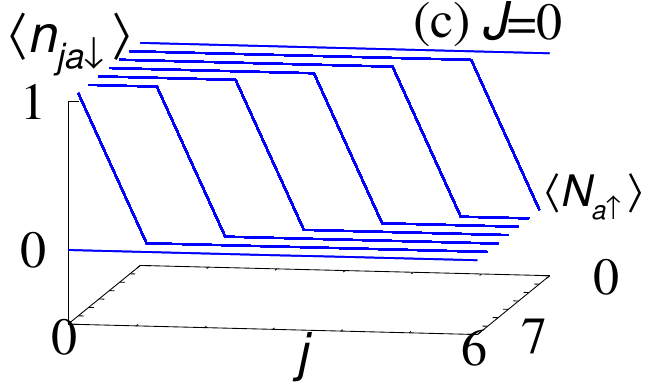}
\end{flushright}
\end{minipage}
\begin{minipage}{0.48\hsize}
\begin{flushleft}
\includegraphics[width=0.5\hsize,clip]{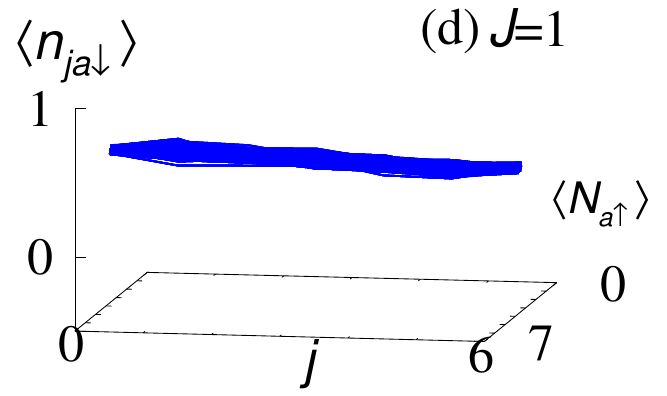}
\end{flushleft}
\end{minipage}
\caption{
(a) and (b): Spectrum of the many-body Hamiltonian for $J=V=0$ and $J=V=1$, respectively.
Blue (red) symbols denote data obtained under the periodic (open) boundary condition.
(c) and (d): Expectation values $\langle \hat{n}_{ja\downarrow}\rangle={}_{\mathrm R}\langle \Phi_n| \hat{n}_{ja\downarrow} |\Phi_n\rangle_{\mathrm R}$ against $j$ and $\langle \hat{N}_{a\uparrow}\rangle ={}_{\mathrm R}\langle \Phi_n| \hat{N}_{a\uparrow} |\Phi_n\rangle_{\mathrm R} $ for $J=V=0$ and $J=V=1$, respectively.
These data are obtained for $(L,t)=(7,1)$ and the subsector $(N,P)=(9,-1)$.
}
\label{fig: HN_N9P-1_OBCPBC}
\end{figure}

Finally, we discuss the case for $(N,P)=(9,-1)$ where orbital $a$ is half-filled.
Figure~\ref{fig: HN_N9P-1_wind} indicates that the many-body Hamiltonian is topologically trivial, which results in fragility of the skin effect at the non-interacting level as discussed in the above.
%
Namely, while the topology of the one-body Hamiltonian induces the extreme sensitivity of the energy spectrum and the expectation values $\langle \hat{n}_{ja\downarrow}\rangle$ to the boundary conditions, such extreme sensitivity is not observed in the interacting case [see Fig.~\ref{fig: HN_N9P-1_OBCPBC}].
%

\end{document}